\documentclass[letterpaper, 10 pt, conference]{ieeeconf}
\usepackage{cite}
\usepackage{amsmath,amssymb,amsfonts}
\usepackage{algorithmic}
\usepackage{graphicx}
\usepackage{textcomp}
\usepackage[]{todonotes}
\usepackage{hyperref}
\usepackage{float}

\title{\LARGE \bf
Online Parameter Estimation using Physics-Informed Deep Learning for Vehicle Stability Algorithms
}

\author{Kemal Koysuren, Ahmet Faruk Keles, Melih Cakmakci}

\begin{document}

\maketitle
\thispagestyle{empty}
\pagestyle{empty}

\begin{abstract}

Physics-informed deep learning is a popular trend in the modeling and control of dynamical systems. This paper presents a novel method for rapid online identification of vehicle cornering stiffness coefficient, a crucial parameter in vehicle stability control models and control algorithms. The new method enables designers to rapidly identify the vehicle front and rear cornering stiffness parameters so that the controller reference gains can be re-adjusted under varying road and vehicle conditions to improve the reference tracking performance of the control system during operation. The proposed method based on vehicle model-based deep learning is compared to other alternatives such as traditional neural network training and identification, and Pacejka model estimation with regression. Our initial findings show that, in comparison to these classical methods, high fidelity estimations can be done with much smaller data sets simple enough to be obtained from a lane-changing or vehicle overtake maneuver. In order to conduct experiments, and collect sensor data, a custom-built 1:8 scaled test vehicle platform is used real-time wireless networking capabilities. The proposed method is applicable to predict derived vehicle parameters such as the understeering coefficient so it can be used in parallel with conventional MIMO controllers. Our $H_{\infty}$ yaw rate regulation controller test results show that the reference gains updated with the proposed online estimation method improve the tracking performance in both simulations and vehicle experiments.

\end{abstract}

\section{INTRODUCTION}
\label{sec:introduction}

Many modern automatic control applications require parameter identification of the plant model. These identified parameters can be used for the initial controller design, online adjustments and monitoring of the health of the system. Recent developments in data gathering and processing in embedded control systems enable combining modeling and control applications with learning techniques. However, due to the computational complexity, control-oriented neural networks are still difficult to implement in many applications\cite{COL}. Additionally, noisy sensor data also make complications to process collected data in a practicable way. To overcome this problem researchers gravitate to the physics-informed deep learning method\cite{PIDL}. There are numerous examples of modeling the plant with learning procedures to increase the fidelity of the control systems. In \cite{MPC}, in order to push the model predictive control's (MPC) performance to the limits, the learning procedure for modeling is improved in a way that the prediction model ensures the best closed-loop performance. In \cite{LSI}, the performance of data-driven modeling and physical modeling approaches for vehicle lateral-longitudinal dynamics are compared. Results show that the data-driven model bests both linear and non-linear physical models. In \cite{RL}, an online model-based reinforcement learning method is proposed to identify vehicle linear tire parameters to maximize maneuverability under variable road conditions such as unknown terrain. 

Automotive control applications are one of those fields where the estimation of the vehicle parameters can improve the performance of the vehicle immensely. Automotive systems contain many complex dynamic mechanisms, many of which are considered lumped parameter models during the control system design phase. For example, in vehicle lateral stability control, the LTI state-space model is widely used. In this model, the cornering coefficient is the hardest parameter to be estimated. It requires experimental data, such as tire lateral forces, which requires an estimator(addition to the IMU sensor data) or additional expensive force sensors. Pacejka's method \cite{pacejka} can estimate the nonlinear behavior of tire models but the method requires many data points and an estimator or additional sensors.

In deep learning, a high number of data points and data sets with labeled data are required to train a network with high accuracy. To train such a network, extensive experiments with many sensors are required.
Another option is to use a so-called physics-informed parameter learning scheme. This type of deep network contains the system physics in the loss function which enables them to be utilized with very few data sets and a lower number of data points. This allows them to be used and trained for a single system with high accuracy. 

The main contribution of this paper is the identification of the vehicle cornering coefficients with a physics-informed learning algorithm fed directly by raw sensor data (i.e. vehicle lateral acceleration, $a_y$ and yaw rate, $r$). The proposed model-based approach requires much smaller data sets to yield lower error values (i.e. better prediction) compared to conventional NN methods and Pacejka's model-based estimations in literature. Hence, the system is suitable to apply as part of a real-time control system.
The rest of the paper is outlined as follows: In Section II the vehicle mathematical model used in the derivation of the physics-based algorithm and in simulations and our scaled vehicle prototype and the experimental setup are introduced. Section III explains the prediction of tire properties with a more conventional method given by Pacejka. In Section IV our physics-informed parameter learning method is presented and compared to traditional estimation with neural networks without the model information. In Section V, first the model estimation accuracy of the physics-informed learning method and conventional deep learning method are compared, then the proposed learning method is implemented on $H_{\infty}$ vehicle yaw-rate regulation algorithm to show its effectiveness with experiments.

\section{VEHICLE MODEL AND TEST SETUP}
\label{sec:II}

\subsection{Mathematical Model} \label{sec:mathmod}
\begin{figure}[h!]
\centerline{\includegraphics[width=(\columnwidth)*5/8]{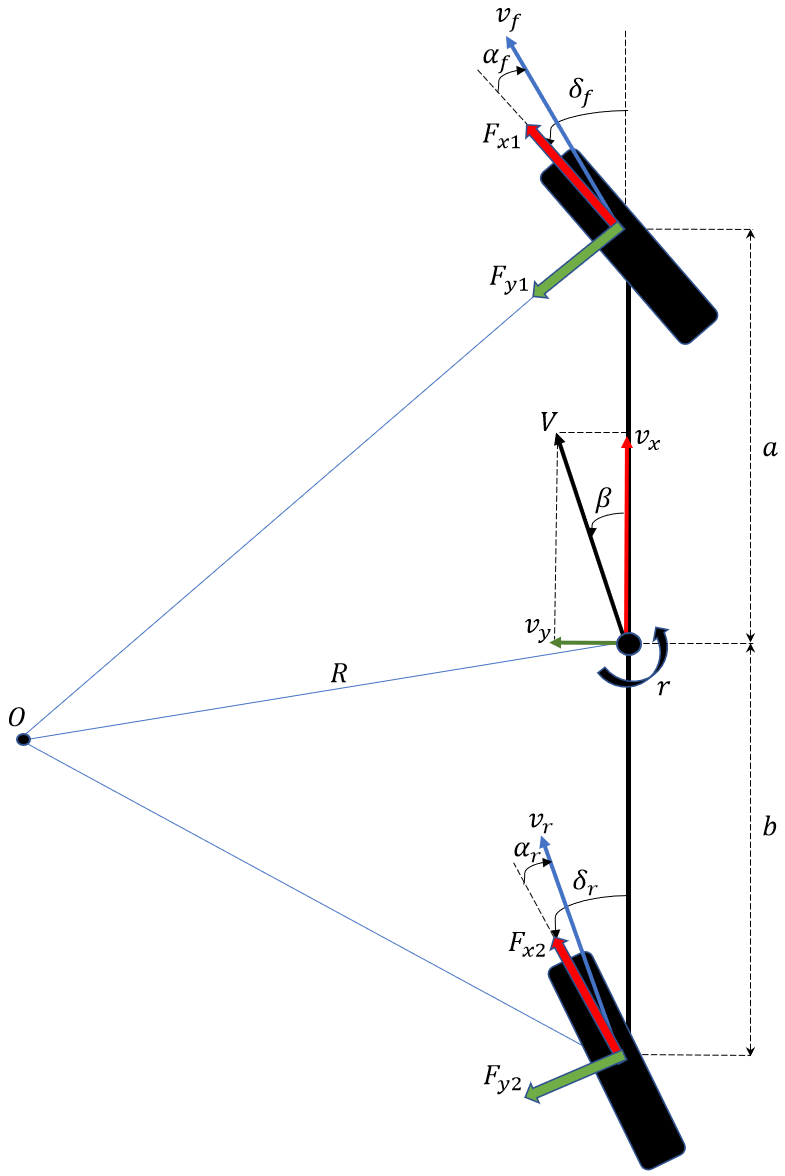}}
    \caption{Vehicle Lateral Model Parameters with Single Track (Bicycle) Assumption.}
    \label{fig:vehlatmod}
\end{figure}

Figure~\ref{fig:vehlatmod} represents single-track 2-DoF vehicle model. It is extensively used in vehicle lateral stability control applications because of its simplicity and accuracy for the majority of the control-oriented problems \cite{acs}. In this model, $\alpha_f$ and $\alpha_r$ are front and rear tire slip angles respectively. $a$ and $b$ are distances between the front wheel to center-of-mass and the rear wheel to center-of-mass. $\delta_f$ and $\delta_r$ represent front and rear wheel steering angles. $v_x$ and $v_y$ are the vehicle's center of mass velocities in x and y directions in local coordinates. $V$ is the vector summation of them and $\beta$ is their inverse tangent. $R$ is the turning radius and $O$ is the center of rotation. $r$ is the yaw rate of the center of mass of the vehicle and $F_{ij}$ are forces acting on the front and rear wheels where $i={x,y}$ and $j={1,2}$. Linearization of this model is required to design such controllers. Hence, by assuming small tire slip angles we obtain a state-space form from \eqref{eq:statespace}-\eqref{eq:B}. Cornering coefficient ($C_{af}$ and $C_{ar}$) estimation is the biggest challenge throughout the linearization process since it requires experiments with vehicle and post-processing (for example curve fitting and finding the slope of the linear region in Pacejka) the collected data to identify those parameters.

\begin{equation}
\label{eq:statespace}
  \begin{aligned}
    \dot{\bf{x}} & = \textbf{Ax+Bu}
  \end{aligned}
\end{equation}

\begin{equation}
\textbf{x}^T =
\begin{bmatrix}
\label{eq:states}
v_y & r
\end{bmatrix}
\end{equation}

\begin{equation}
\textbf{u}^T =
\begin{bmatrix}
\label{eq:inputs}
\delta _1 & \delta _2
\end{bmatrix}
\end{equation}

\begin{equation}
\textbf{A} =
\begin{bmatrix}
\label{eq:A}
-\frac{C_{af}+C_{ar}}{mV_0} & \frac{bC_{ar}-aC_{af}}{mV_0}-V_0 \\
\frac{bC_{ar}-aC_{af}}{I_zV_0} & -\frac{a^2C_{af}+b^2C_{ar}}{I_zV_0} 
\end{bmatrix}
\end{equation}

\begin{equation}
\textbf{B} =
\begin{bmatrix}
\label{eq:B}
\frac{C_{af}}{m} & \frac{C_{ar}}{m}\\
\frac{aC_{af}}{I_z} & -\frac{bC_{ar}}{I_z}
\end{bmatrix}
\end{equation}

\begin{figure}[h]
\centerline{\includegraphics[width=(\columnwidth)]{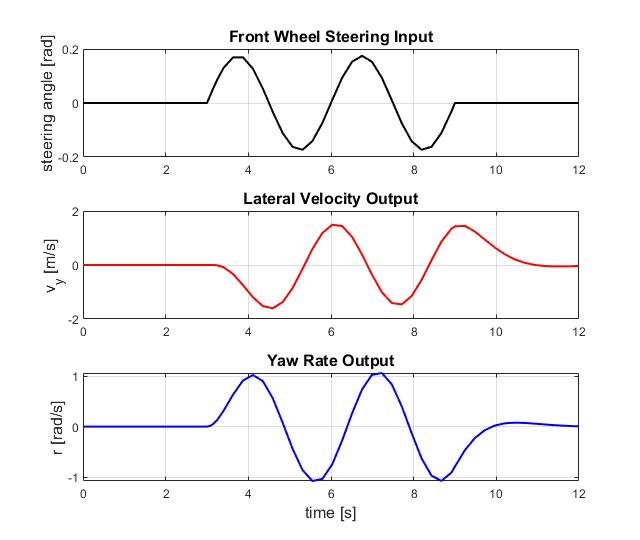}}
    \caption{Typical Simulation Output}
    \label{fig:typsimout}
\end{figure}

Figure~\ref{fig:typsimout} is an example of the response of our linear vehicle model to sinusoidal steering input. The accuracy of the data collected from the sensors ($v_y$ and $r$) will be compared with this graph.

\subsection{Test Vehicle and Experiments} \label{sec:setup}

\begin{figure}[!t]
\centerline{\includegraphics[width=(\columnwidth)]{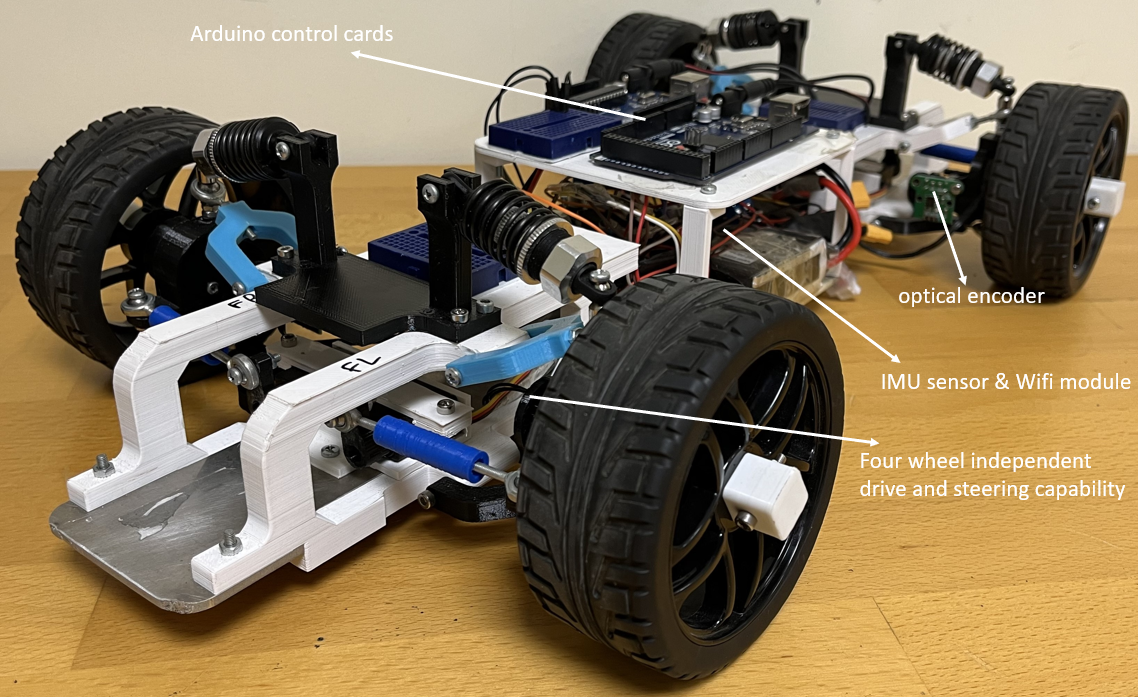}}
    \caption{Scaled Test Vehicle with 4WD-4WS equipped with inertial measurement sensing and wireless communication capabilities for experiments.}
    \label{fig:testvehicle}
\end{figure}

In order to conduct experiments and collect sensor data to feed the physics-informed learning algorithm, we assembled a custom-built 1:8 scaled test vehicle (Figure~\ref{fig:testvehicle}) with four-wheel-steering and four in-wheel electric motor independent drive features. There are two Arduino-based control cards on the vehicle. While the first board collects the sensor data, runs feedback control algorithms and communicates with the Matlab server, with a WiFi module, where learning algorithms are run, the second board provides the motor actuation commands. Each wheel has an optical encoder sensor. Thus, we can measure the wheel rotation speed of the vehicle and the speed of the center of gravity with high accuracy. There is also a 9-axis IMU sensor (BNO055) mounted on the center of the gravity of the vehicle. Parameters of test vehicle are: $m=2.15[kg]$, $I_z=0.085[kgm^2]$, $a=b=0.17[m]$. Lateral acceleration and yaw rate values are exported from the IMU sensor. From the yaw rate and lateral acceleration $\Dot{v_y}$ is calculated as:
\begin{equation}
    \Dot{v}_y=a_y-v_xr
\end{equation}

\section{PACEJKA's TIRE MODELLING}
\label{sec:pacejka}

Before moving to the proposed parameter estimation method, the results of the most common method called Pacejka tire model (equation \ref{eq:Fy} and \ref{eq:phi}), or Magic tire formula, from \cite{pacejka} that is used for cornering coefficient estimation will be analyzed. 

\begin{equation}
    F_y = Dsin(Carctan(B\Phi))
    \label{eq:Fy}
\end{equation}
\begin{equation}
    \Phi = (1-E)\alpha+(E/B)arctan(B\alpha)
    \label{eq:phi}
\end{equation}

In these equations, $D=max(F_y)$ and $C=1.30$, $B$ and $E$ are unknown. Using the same sets of data collected for learning, 2 unknown parameters in equation \eqref{eq:Fy}-\eqref{eq:phi} are calculated with the least square curve fitting. The collected data needs to be pre-processed to calculate tire lateral force and tire slip angle. For lateral tire force estimation, sliding mode observer from  \cite{SMO} is applied. For front and rear tire slip angle calculation equation \eqref{eq:alphaf}-\eqref{eq:alphar} are used:

\begin{equation}
    \alpha_f = \delta_f-\frac{v_y-ar}{v_x}
    \label{eq:alphaf}
\end{equation}
\begin{equation}
    \alpha_r = -\frac{v_y-br}{v_x}
    \label{eq:alphar}
\end{equation}

\begin{figure}[h]
\centerline{\includegraphics[width=(\columnwidth)]{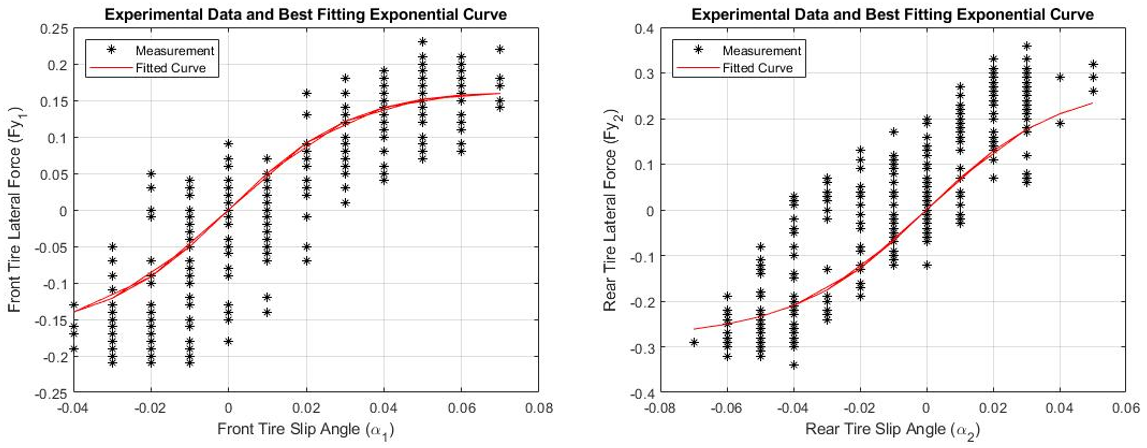}}
    \caption{Tire Model Regression}
    \label{fig:pacejka}
\end{figure}

Best fitting exponential curves for front and rear tires are shown on Figure~\ref{fig:pacejka}.

Slope of the linear region of these curves are $C_{af}=4.59$ and $C_{ar}=6.45$ respectively. Since the number of collected data points is insufficient and quality of the set of the data is low due to the noise of the sensors, the error value according to the equation in \eqref{eq:error} can be calculated as 3.62, its significance will be discussed in the next section.

\section{PHYSICS-INFORMED PARAMETER LEARNING}
\label{sec:PIDL}

The goal of the deep learning algorithms in this system is to estimate the $C_{af}$ and $C_{ar}$ values from the input data, which are $r$, $\Dot{r}$, $v_y$, $\Dot{v_y}$, $\delta_1$, $\delta_2$, $v_x$. In order to find an estimation, a simple deep neural network is created to take advantage of automatic differentiation and gradient descent. The network consists of 6 layers,which are an input layer, 3 fully connected layers with 20 neurons, 1 fully connected layer with 2 neurons and an activation function with the form of:
\begin{equation}
    Z=Z_{mean}*(1+Z_{range}*tanh(X))
\end{equation}

The gradient for the learnable parameters L for back-propagation becomes:

\begin{equation}
    \dfrac{\partial L}{\partial X}=Z_{mean}*Z_{range}*\dfrac{\partial L}{\partial Z}*(1-tanh(X)^2)
\end{equation}

As $X$ moves from -$\infty$ to $\infty$, $Z$ moves from $Z_{mean}*(1-Z_{range})$ to $Z_{mean}*(1+Z_{range})$. Taking the output of this layer as $C_{af},C_{ar}$ the loss function is constructed from the governing differential equation as:

\begin{equation}
    Loss=\dfrac{1}{2N}*\sum^N_{i=1}\bf{\alpha}*(\Dot{\bf{x}}-\bf{A}\bf{x}-\bf{B}\bf{u})^T(\Dot{\bf{x}}-\bf{A}\bf{x}-\bf{B}\bf{u})
    \label{eq:loss}
\end{equation}

where $\alpha$ is a 2-element constant vector that is used for scaling between equations. $\bf{A}$, $\bf{B}$, $\bf{x}$, $\bf{u}$ are the state space parameters from mathematical model in Section 2. An additional loss function is utilized to ensure that $C_{af}$ and $C_{ar}$ values are constant in time by taking the mean value of $C_{af}$, $C_{ar}$ and comparing them to the value of $C_{af}$, $C_{ar}$ at each time step. With this setup, four different experiment results are examined. The deep learning setup is outlined in Figure \ref{fig:my_label}(a). $Z_{mean}$ is taken as 10 and $Z_{range}$ is taken as 90\%. The network is updated with an adaptive moment estimation method with a learning rate taken as 0.001 with a decay of 0.0005 \cite{b3}.
\begin{figure}[h]
    \centering
\includegraphics[trim=25 20 35 10,clip, width=(\columnwidth)]{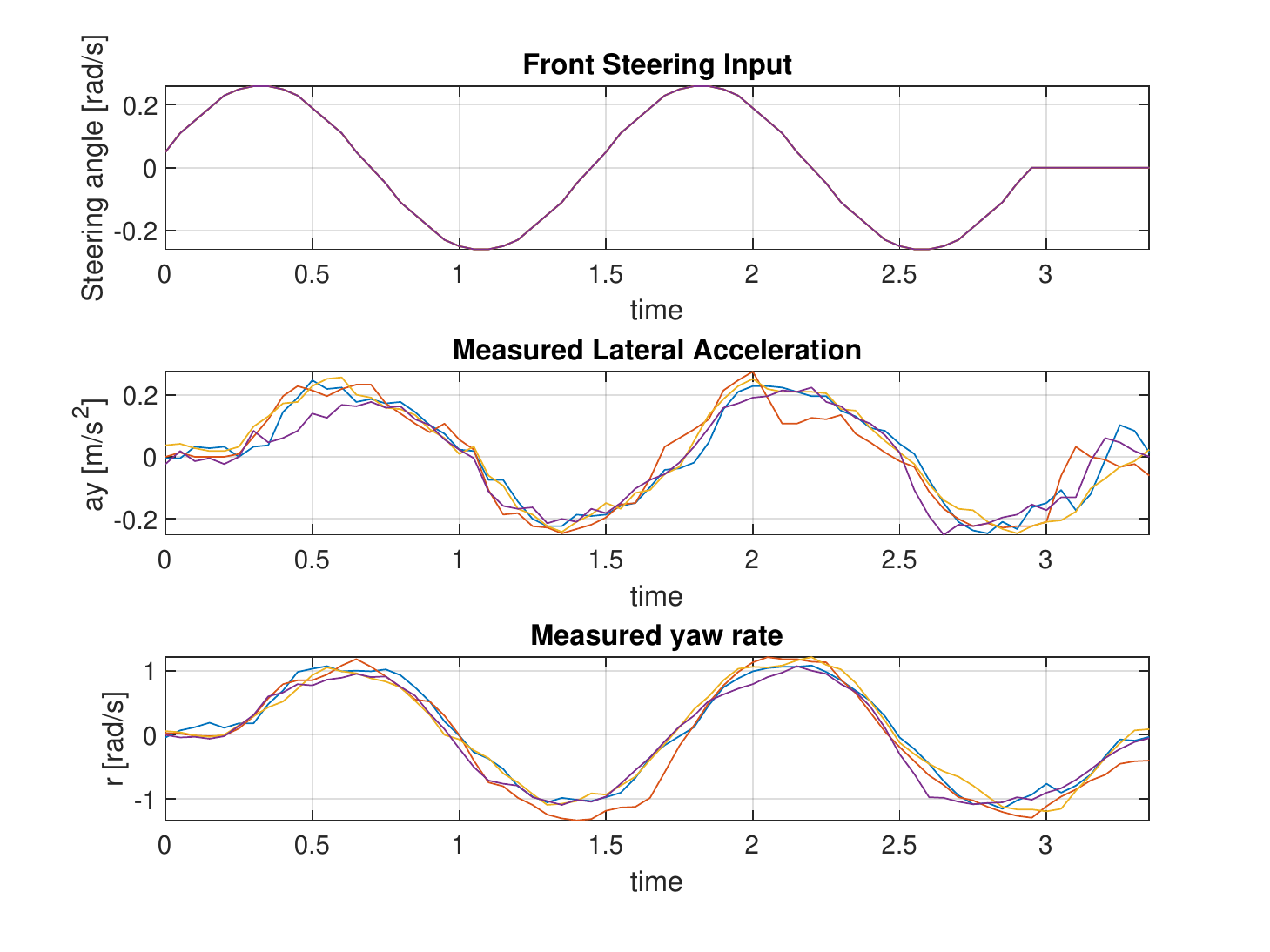}
    \caption{Results of Experiments}
    \label{fig:my_label}
\end{figure}
From the experiment results, $C_{af}$ and $C_{ar}$ are:

\begin{equation*}
    \begin{bmatrix}
    C_{af}\\
    C_{ar}
    \end{bmatrix}=\begin{bmatrix}
    7.26&8.67&8.59&8.05\\
    9.83&9.45&10.35&9.21
    \end{bmatrix}
\end{equation*} 

From four separate time-series trajectories of experimental data, $C_{af}$ can be calculated as 8.14 with 8.65\% relative uncertainty and $C_{ar}$ can be calculated as $9.71$ with $5.87\%$ relative uncertainty. To calculate the error in this parameter prediction, experiment results are compared with simulations carried out with predicted cornering stiffness. Because the data involves 0 values, percent error cannot be used. The error is formulated as:

\begin{equation}
    e=\int(|v-v_{simulated}|+|r-r_{simulated}|)dt
    \label{eq:error}
\end{equation}

\begin{figure}[h]
    \centering
    \includegraphics[width=\columnwidth]{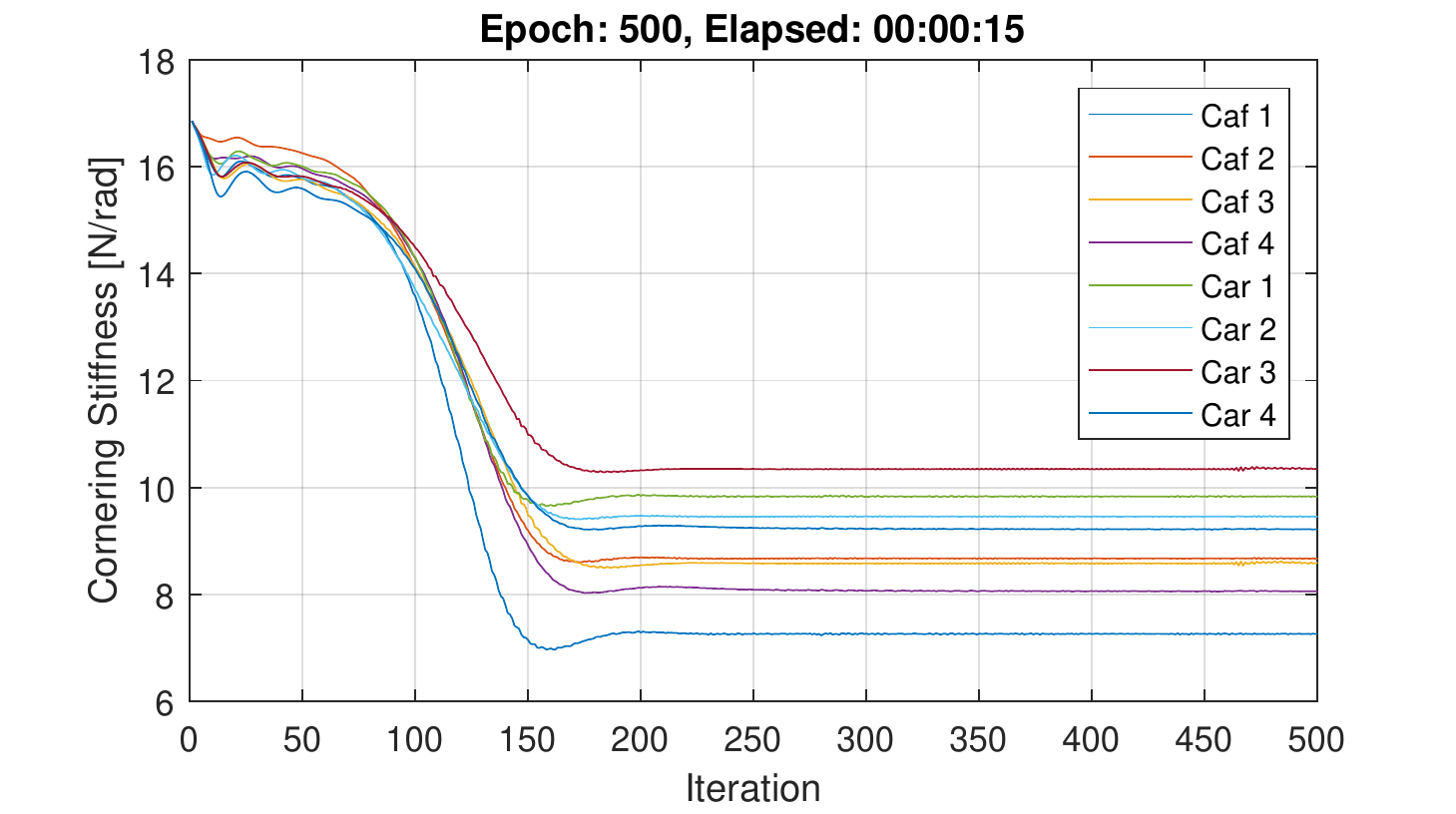}
    \caption{Convergence results with physics informed deep learning for  4 experiments}
    \label{fig:Conv}
\end{figure}

The error values for these results are $2.09$,$2.81$,$2.18$,$3.06$. The convergence plot for the first 4 results is given in Figure \ref{fig:Conv}.

The advantage of physics-informed neural networks to the traditional networks is that the physics-based approach can work with single data set rather than a large batch of labeled data in a much shorter amount of time which enables real-time online training \cite{b4}. Moreover, requiring a large batch of labeled data makes using traditional neural networks with experimental works challenging. Simulation results can be used to train traditional neural networks for predicting results from experimental data however, sensor noise and drift reduce the accuracy of the results. To demonstrate this, a traditional regression neural network that consists of a BiLSTM layer, 2 fully connected layers and the custom output layer described above, is trained (using the procedure outlined in Figure \ref{fig:my_label}(b)) by using simulation results with $C_{af}$ and $C_{ar}$ values ranging from 1 to 19 with 1 interval. The results are:

\begin{equation*}
    \begin{bmatrix}
    C_{af}\\
    C_{ar}
    \end{bmatrix}=\begin{bmatrix}
    1.95&1.91&1.95&1.97\\
    5.41&4.89&5.23&5.38
    \end{bmatrix}
\end{equation*}

\begin{figure}[h]
    \centering
    \includegraphics[height=5cm,width=\columnwidth]{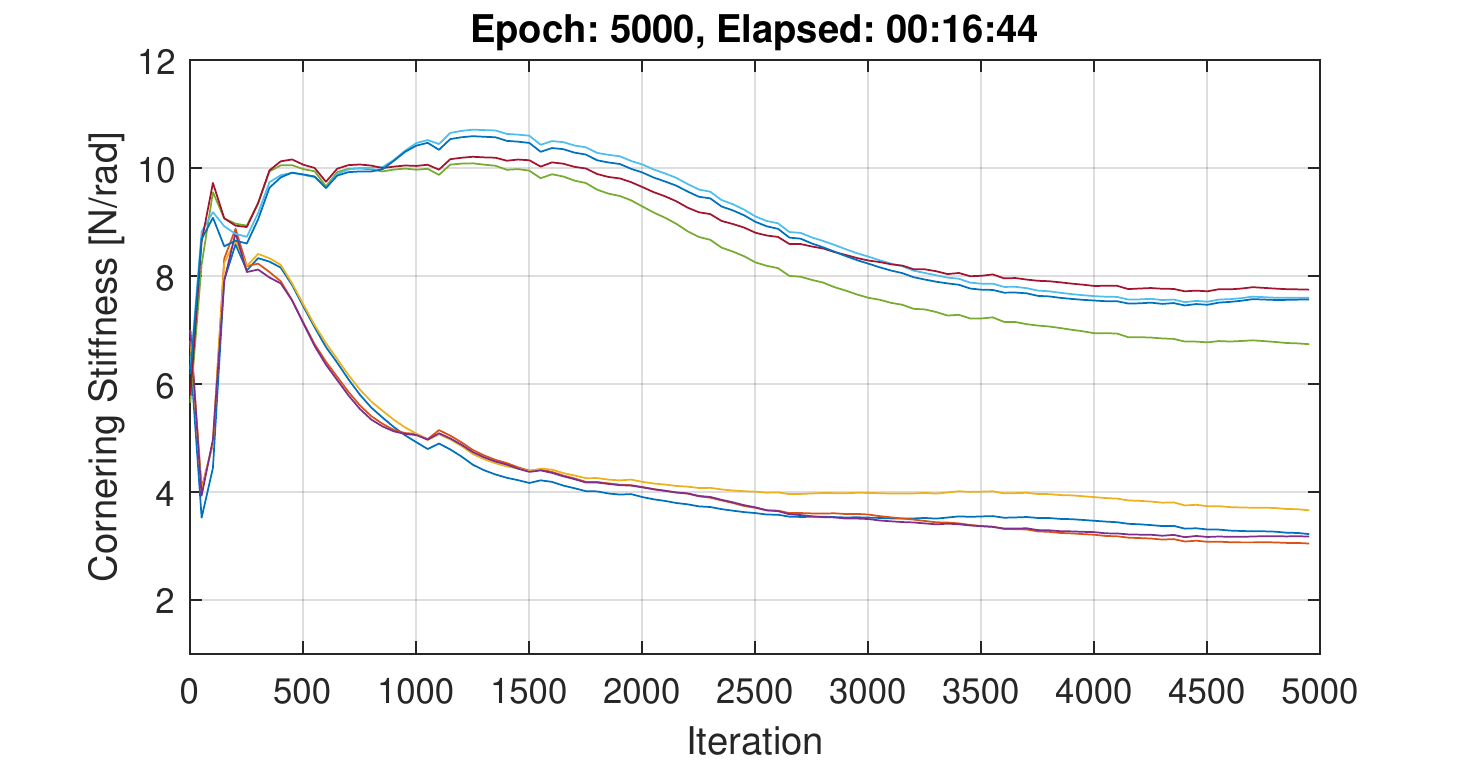}
    \caption{Convergence results with regression deep learning for 4 experiments}
    \label{fig:Conv2}
\end{figure}

The error values for these results are 3.25,4.12,3.77,4.03. From the loss values, it can be seen that the results from the traditional neural network are sub-optimal (i.e. not accurate estimations) using the same experimental datasets. These loss values may be further improved with more extensive data collection but the trend on the data seems to be sub-optimal in any case. The convergence plot for the 4 results is given in Figure \ref{fig:Conv2}.

\begin{figure}[h]
    \centering
    \includegraphics[trim=50 20 50 10,clip,width=(\columnwidth)]{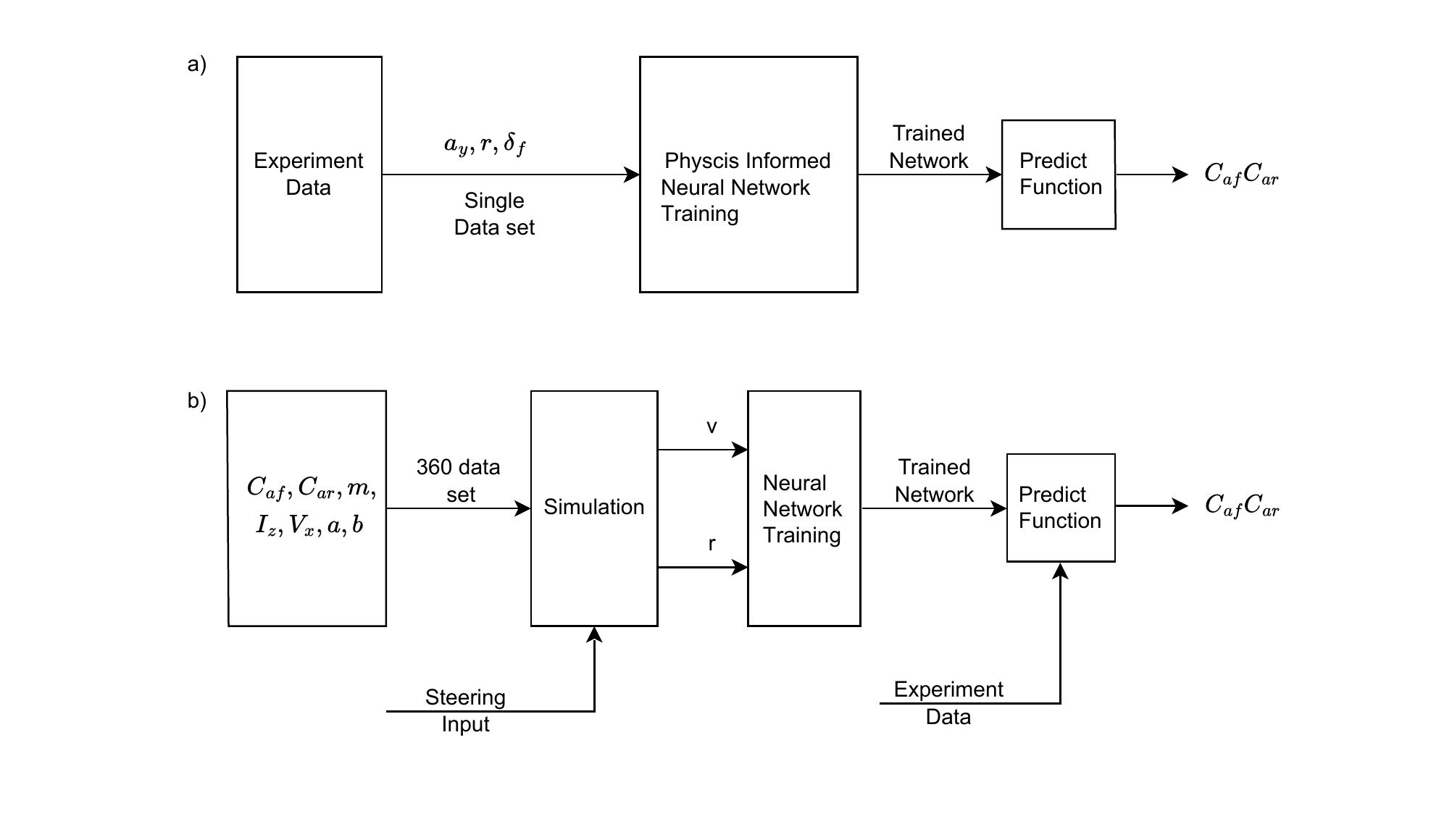}
    \caption{a) Physics Informed Training b)Traditional Regression Training}
    \label{fig:my_label}
\end{figure}

\section{CONTROLLER IMPLEMENTATION WITH ONLINE PARAMETER ESTIMATION AND EXPERIMENTS}

In this section, we investigate how online estimation can be beneficial to control applications using a MIMO $H_{\infty}$ controller design example with variable reference signal generation. Figure~\ref{fig:onlinecontrol} shows the block diagram of the closed system. Our  proposed controller has two primary parts: a reference generator, and a MIMO controller. The MIMO controller is designed such that the system is stable to changes in the cornering stiffness tracking performance. The reference generator uses cornering stiffness values to generate the current MIMO controller reference. The parameter values it uses to calculate these commands are updated online estimation from the model-based deep learning algorithm as discussed earlier.

\subsection{Reference Generator}
The driver initiates the reference steering and velocity commands. In reference generator, with driver commands, the desired lateral acceleration gain and yaw rate gain are generated using \eqref{eq:rref} and \eqref{eq:ayref} as mentioned in \cite{ACC23}.

\begin{equation}
\label{eq:rref}
    r_{ref}=\delta_{1,in}\dfrac{v_{x}}{(a+b)+K_{us}v_{x}^2}
\end{equation}

\begin{equation}
\label{eq:ayref}
    a_{y,ref}=\delta_{1,in}\dfrac{v_{x}^2}{(a+b)+K_{us}v_{x}^2}
\end{equation}
where understeer coefficient, $K_{us}$, is given as
\begin{equation}
\label{eq:Kus}
    K_{us}=\dfrac{mb}{(a+b)C_{af}}-\dfrac{ma}{(a+b)C_{ar}}
\end{equation}

The learning algorithm updates $C_{af}$ and $C_{ar}$ values according to the data received from the vehicle and keeps the understeering coefficient updated in \eqref{eq:Kus} so that the reference generator block calculates its outputs based on the latest driving conditions.
\subsection{Optimal MIMO Controller}
\begin{figure*}[h!]
\centering
\includegraphics[width=(\columnwidth*10/6)]{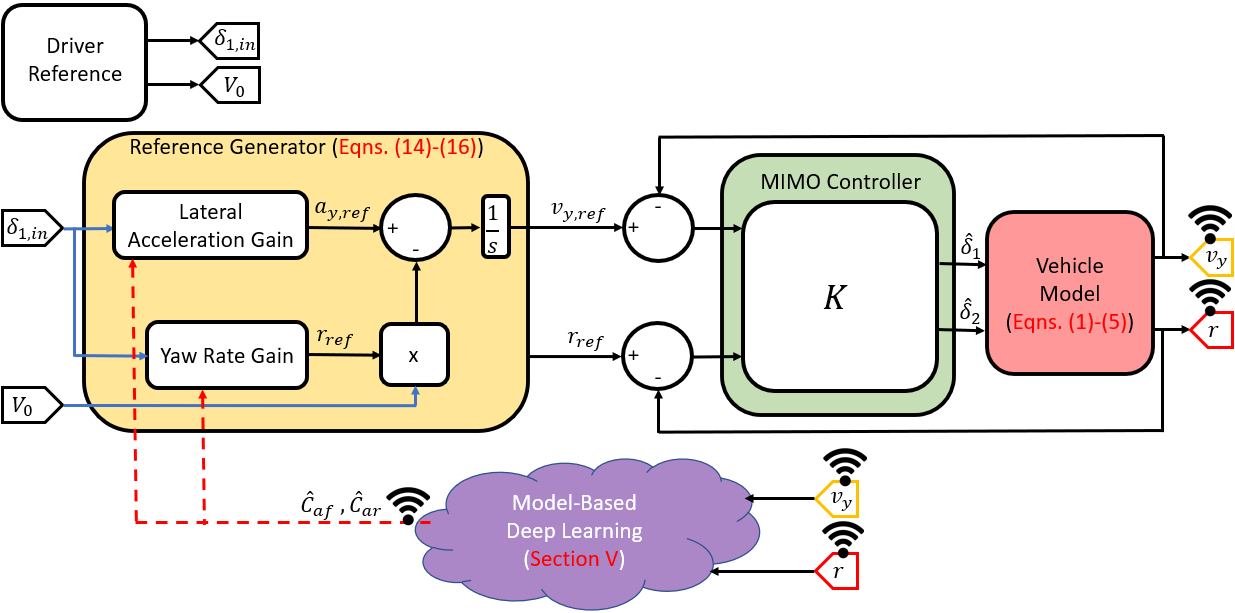}
    \caption{Feedback Control with Updated References Block Diagram}
    \label{fig:onlinecontrol}
\end{figure*}

The optimal MIMO controller shown in Figure \ref{fig:onlinecontrol} regulates front and rear steering angles using state error signals, to improve reference tracking. Both states from the vehicle model are used as feedback for the controller. The MIMO $H_{\infty}$ controllers are designed using mixed-sensitivity minimization with the methods available from sources such as \cite{zhou1996robust} and applicable for a range of cornering stiffness parameters. Input of the controller is set as yaw rate($r$) and control outputs are set as front($\delta_{1}$) and rear($\delta_{2}$) steering. 

\subsection{Comparison of PIDL and RDL Parameter Estimation Performance} \label{sec:benchmark}

In this section, using the state-space model mentioned in Section \ref{sec:mathmod}, two  MIMO $H_{\infty}$ controllers are designed using the cornering stiffness parameters estimated with physics-informed deep learning (PIDL) and regression deep learning (RDL) methods. These controllers are then implemented on the test vehicle (Section \ref{sec:setup}) and a typical lane change maneuver test given Figure \ref{fig:typsimout} is performed to obtain and compare the error values of estimation results of both learning methods from Section \ref{sec:PIDL}. Cornering coefficients estimation of PIDL are taken as $C_{af}=8.14$ and $C_{ar}=9.71$, and coefficients from RDL are taken as $C_{af}=1.95$ and $C_{ar}=5.23$ (i.e. mean of the estimation set).

\begin{figure}[h]
\centering
\includegraphics[width=(\columnwidth)*10/11]{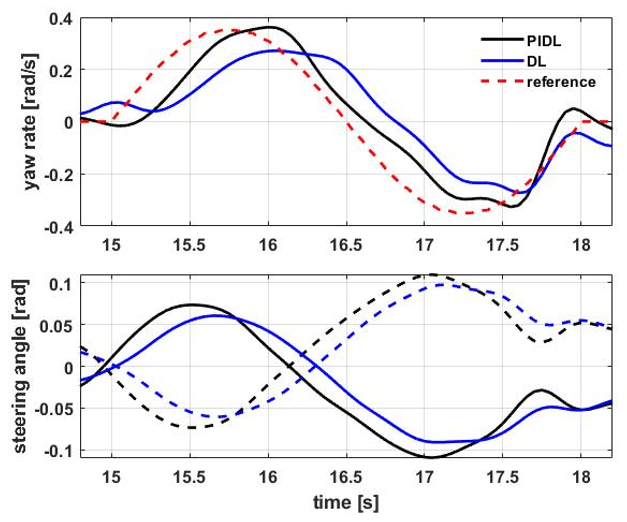}
    \caption{Controller Performances (PIDL: physics-informed deep learning, DL: regression deep learning). In bottom figure dotted lines: rear wheel steering angle and continuous lines: front wheel steering angle}
    \label{fig:benchmark}
\end{figure}

Figure~\ref{fig:benchmark} presents the result of the feedback control experiments on the test vehicle. It shows that the yaw rate reference tracking performances of both controllers. The regulator designed with PIDL cornering parameters has better yaw reference tracking than the RDL. This verifies that PIDL estimation has better modeling accuracy since the controllers designed using this model generate better performance.

\subsection{PIDL Experiments with Varying Drive Conditions}
The discussion in the previous section show that prediction generated by the PIDL algorithm is more accurate even for short burst of data that can be obtained during a lane change maneuver. Therefore, it is also possible to use this method to update vehicle controller parameters online to improve performance. 

As an example scenario we start with the conditions such that the test vehicle cruising with constant longitudinal velocity $v_x=1.2[m/s]$ and the vehicle controller operating with the parameters $C_{af}=8.14$ and $C_{ar}=9.71$ as calculated earlier from the experiment data. When the vehicle moves to another surface where road-tire physical interaction is much higher, the cornering stiffness coefficients will change.  The optimal MIMO controller can be designed such changes in mind but using the older coefficients in the reference generator will generate a sub-optimal overall controller since the understeering coefficient of the vehicle is changed in the new conditions.


\begin{figure}[h]
\centerline{\includegraphics[width=(\columnwidth)*10/11]{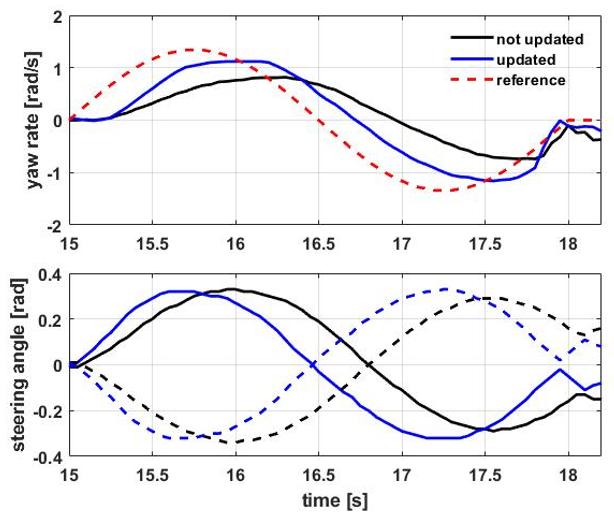}}
    \caption{Yaw Rate Comparison Experiment (not updated: constant $K_{us}$ system, updated: updated $K_{us}$ parameter with online estimation). In bottom figure dotted lines: rear wheel steering angle and continuous lines: front wheel steering angle}
    \label{fig:yawrate}
\end{figure}

Figure~\ref{fig:yawrate} shows the results of a such experimental scenario where from the from proposed physics informed deep learning algorithm, the new coefficient values are calculated as $C_{af}=2.36$ and $C_{ar}=4.38$. It shows that the control structure without $K_{us}$ reference update deviates from the yaw rate reference and becomes less responsive to the inputs because of higher road-tire interaction that was not compensated for in the outdated reference generator. On the other hand, the control structure with an updated reference value, i.e. online parameter estimation system, is not affected by the road condition change much as its yaw rate tracking capability is better.

\section{CONCLUSIONS}

This paper introduces a novel physics-informed learning algorithm to estimate rapid vehicle tire cornering coefficients suitable to be used in conventional lateral stability control algorithms. Our simulation and experimental results show that, compared to the conventional regression-based learning algorithm and widespread Pacejka's method, the proposed method requires less data-set to fit collected data points. Hence, it allows real-time online parameter identification. The predicted coefficients straight forward and accurate enough to be used reference gain updates to improve the performance of control algorithms under varying road surfaces. The performance of online estimation, with experiments on the test vehicle, is verified by implementing it on an $H_\infty$ yaw-rate regulator. Future work includes the extension of the method to more detailed models so that more parameters can be estimated with smaller data sets.


\end{document}